\documentclass[12pt]{article}

\usepackage{epsfig}

{\end{list}}
\newcounter{enumct}

\newcommand{\captive}[1]{\rule{5mm}{0mm}%
\begin{minipage}{150mm}\caption[small]{#1}\end{minipage}}

\newcommand{\beq}{\begin{equation}}
\newcommand{\eeq}{\end{equation}}
\newcommand{\beqa}{\begin{eqnarray}}
\newcommand{\eeqa}{\end{eqnarray}}

\newcommand{\lsim}{\buildrel < \over {_\sim}}
\newcommand{\half}{\frac{1}{2}}

\newcommand{\ie}{{\it i.e.}}
\newcommand{\eg}{{\it e.g.}}
\newcommand{\cf}{{\it cf.}}
\newcommand{\etal}{{\it et al.}}
\newcommand{\gev}{{\rm GeV}}

\newcommand{\morder}[1]{{\cal O}(#1)}
\newcommand{\eq}[1]{Eq.\ (\ref{#1})}

\newcommand{\as}{\alpha_s}

\newcommand{\ket}[1]{\vert{#1}\rangle}

\newcommand{\PL}[3]{Phys.\ Lett.\ {{\bf#1}} ({#3}) {#2}}

\newcommand{\PR}[3]{Phys.\ Rev.\  {{\bf#1}} ({#3}) {#2}}
\newcommand{\PRL}[3]{Phys.\ Rev.\ Lett.\ {{\bf#1}} ({#3}) {#2}}

\begin{document}
 
\sloppy

\pagestyle{empty}

\begin{flushright}
        NORDITA-1999/76 HE\\
        hep-ph/9911486\\
\end{flushright}

\vskip 1cm

\begin{center}
{\LARGE\bf ELFE Physics\footnote{Talk given at the{\em Workshop on the
development of future linear electron-positron colliders
for particle physics studies and for research using free electon lasers,}
Lund, Sweden, 23-25 September 1999.}}\\[5mm] {\Large Paul Hoyer}
\\[3mm] {\it Nordita}\\[1mm]
{\it DK-2100 Copenhagen, Denmark}\\[1mm]
{\it E-mail: hoyer@nordita.dk}\\[5mm]

{\bf Abstract}\\[1mm]
\begin{minipage}[t]{140mm}
This is an introductory review of central topics in hadron physics that
are addressed by high intensity, continuous electron beam facilities in the
multi-GeV range. Exclusive processes are a crucial tool for increasing
our knowledge of hadron wave functions beyond what can be learned from
studies of hard inclusive processes. Data shows precocious scaling, suggesting
a simple underlying picture. Important conceptual advances in the application
of PQCD is being made.
\end{minipage}\\[5mm]

\rule{160mm}{0.4mm}

\end{center}

\section{General Aims}

The ELFE (Electron Laboratory for Europe) project aims at the construction of a
high intensity, continuous electron beam with good energy resolution in the 15
\ldots 30 GeV energy range \cite{elfe}. In this overview I shall give some of
the physics motivations in a form that should be accessible for persons working
outside this field.

A photon of virtuality $Q^2 \simeq\ 10\ \gev^2$, typical of what can
be reached in the ELFE energy range, has a wavelength $1/Q \sim 
0.06$ fm. ELFE will measure hadron and nuclear wave functions with
this resolution. 

In view of the moderate energy scale, one might ask what new ELFE
can tell us? Nucleon and nuclear structure functions are known \cite{sf} over a
vastly larger range of $x$ and $Q^2$ than can be covered at ELFE. Indeed, the
experimental limit on the radius of the proton constituents, quarks and gluons,
is better than 0.001 fm. Moreover, the data on $F_2(x,Q^2)$ is in impressive
agreement with the predictions of perturbative QCD (PQCD).

The short answer is that there is a lot more to proton structure than is
revealed by the quark and gluon structure functions. To illustrate this,
consider the `Fock expansion' of the proton state $\ket{p}$ in terms of its
quark and gluon constituents, at equal Light-Cone (LC) time $x^+=x^0+x^3$,
\beqa
\ket{p}&=& \int \left[\prod_i\, dx_i\, d^2k_{\perp i}\right] \left\{
\Psi_{uud}(x_i,k_{\perp i},\lambda_i) \ket{uud} \right. \nonumber \\
&+& \left. \Psi_{uudg}(\ldots) \ket{uudg}+ \ldots + \Psi_{\cdots}(\ldots)
\ket{uudq\bar q}+ \ldots \right\} \label{fock}
\eeqa
Each Fock state $\ket{uud\ldots}$ is weighted by an amplitude $\Psi$
which depends on the LC momentum fractions $x_i$ ($\sum_i x_i = 1$), the
relative transverse momenta $k_{\perp i}$ ($\sum_i k_{\perp i} = 0$) and the
helicities $\lambda_i$ of its constituents. A complete description of the
proton is equivalent to specifying all its Fock amplitudes.

The inclusive single parton distributions $F_{j/p}(x,Q^2)$ give the probability
for finding (at resolution $1/Q$) a parton $j$ which carries the momentum
fraction $x$ of the proton. This parton can be in any Fock state of the proton.
The inclusive parton distribution can thus be schematically expressed in terms
of the Fock amplitudes as
\beq
F_{j/p}(x,Q^2)= \sum_n \int^{k_{\perp}^2<Q^2}\left[ \prod_i\, dx_i\,
d^2k_{\perp i}\right] |\Psi_n(x_i,k_{\perp i})|^2 \sum_j \delta(x-x_j) 
\label{strfn}
\eeq
It should be clear from this grand average over Fock states that Deep Inelastic
Scattering (DIS), despite its central role in advancing our understanding of
hadrons and in establishing QCD as the theory of strong interactions, still only
provides a limited glimpse of the structure of the proton. This fact appears
sometimes to be forgotten in review talks on QCD.

Further qualitative progress in our understanding of QCD and hadron structure
can and has been made by considering other processes than hard inclusive
scattering, where factorization theorems allow PQCD to be rigorously applied.
Such processes tend to be difficult to measure, due to small cross sections and
precise specifications of the final state. This is the arena in which ELFE can
make a contribution, thanks to the unique qualities of its electron beam.

In the next section I give some examples of PQCD processes which are
currently under theoretical and experimental study. I then briefly describe the
present options for ELFE and conclude with some general remarks.

\section{PQCD Processes}

\subsection{Factorization in exclusive reactions}

As an example of the kind of question about proton structure that is not
addressed by data on inclusive scattering, consider the following:

\begin{quote}
{\it What is the probability that the proton contains only its three
valence quarks $uud$, in a configuration of transverse size $\sim 0.1$ fm?}
\end{quote}

\noindent
This probability is given by the square of the proton `distribution amplitude'
$\varphi_p$,
\beq
\varphi_p(x_i,Q^2) = \int^{k_{i\perp}^2 < Q^2} \prod_i d^2\vec k_{i\perp}
\Psi_{uud}(x_i,k_{i\perp})  \label{distamp}
\eeq
which is just the valence Fock amplitude of \eq{fock}, integrated over the
relative transverse momenta up to $Q \sim 2$ GeV at fixed LC momentum
fractions $x_i$.

In order to study compact proton Fock states we need to measure hard {\em
exclusive} scattering \cite{bl}, where coherence is required over the whole
proton rather than just over a single quark constituent. For example, the
proton electromagnetic form factor $F_p(Q^2)$, measured in elastic $ep \to ep$
scattering at high momentum transfer $Q^2$, is determined by the distribution
amplitude (\ref{distamp}),
\beq
A(ep\to ep) = \int_0^1 \left[ \prod_{i=1}^3 dx_i dy_i \right] \,
\varphi_p(x_i,Q^2)\, T_H\,\varphi_p(y_i,Q^2) \{1+\morder{1/Q^2}\}  \label{epep}
\eeq
where $T_H$ is a hard scattering amplitude calculable in PQCD (Fig. 1 a). 

\begin{figure}
\begin{center}
\mbox{\epsfig{file=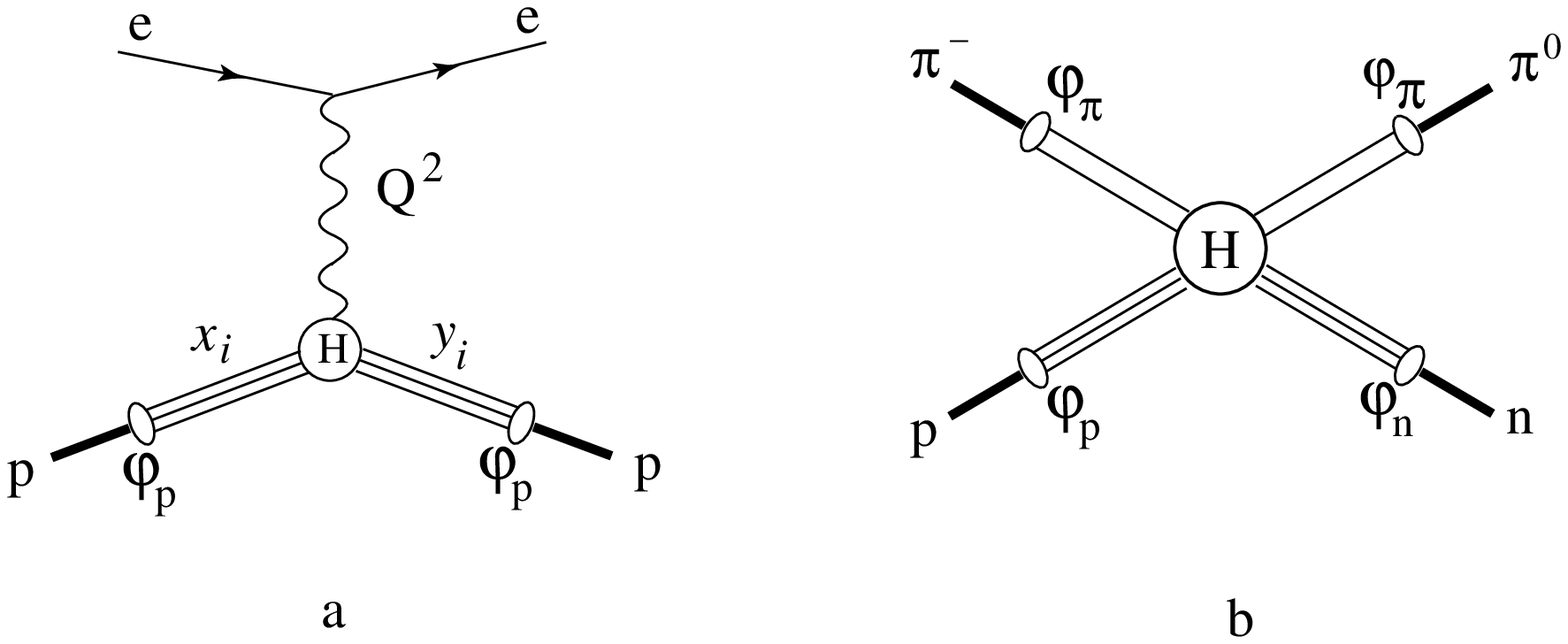,width=160mm}}
\end{center}
\captive{a. Elastic $ep\to ep$ scattering at large $Q^2$ factorizes into a
product of proton distribution amplitudes $\varphi_p$ and a hard electron
scattering from the compact valence Fock state $\ket{uud}$. b. An analogous
factorization is illustrated for the large angle process $\pi^-p \to \pi^0n$.
\label{Fig1_Lund}}
\end{figure}

The factorization of hard exclusive scattering amplitudes into long distance
parts that depend on the hadron wave functions ($\varphi_h$) and a
hard calculable subprocess ($T_H$) is quite general \cite{bl}. As an example,
the factorization of $A(\pi^-p \to \pi^0n)$ at large CM energy and fixed
scattering angle is illustrated in Fig. 1b. The same distribution amplitudes
occur in many hard exclusive processes, allowing cross checks of the PQCD
factorization. The (logarithmic) dependence on the hardness scale $Q^2$ is
predicted, \eg,
\beq
\varphi_p(x_i,Q^2)=120 x_1 x_2 x_3 \delta(1-x_1-x_2-x_3) \sum_{n=0} \left[
\frac{\as(Q^2)}{\as(Q_0^2)} \right]^{\lambda_n} C_n P_n(x_i)
\label{qdep}
\eeq
The anomalous dimensions form an increasing series
\beq
\lambda_0= \frac{2}{27} < \lambda_1= \frac{20}{81} < \lambda_2= \frac{24}{81} <
 \ldots
\eeq
implying that each successive term in \eq{qdep} decreases faster with $Q^2$
than the previous one. The $P_n$ are Appell polynomials, $P_0=1$, 
$P_1=x_1-x_3$, $P_2= 1-3x_2,\ \ldots$ and the $C_n$ are constants which
characterize the proton wave function and have to be determined from
experiment.

The PQCD framework for exclusive processes is in many ways analogous to that
of inclusive reactions. The distribution amplitude takes the place of parton
distributions, with the `scaling violations' specified by \eq{qdep}. The good
news is that the asymptotic shape of the distribution amplitude in the $Q^2 \to
\infty$ limit is predicted, \eg, $\varphi_p^{as}(x_i) \propto x_1 x_2 x_3
\delta(1-x_1-x_2-x_3)$. The bad news is that an exclusive process typically
measures only the absolute square of the convolution of several distribution
amplitudes with a perturbative subprocess amplitude $T_H$, \cf\ \eq{epep}.
Moreover, exclusive cross sections are small and difficult to measure -- this
is why facilities like ELFE are called for.

\subsection{Tests of scaling}

A basic test of the PQCD factorization of exclusive processes is that the
cross section has the predicted power dependence on the hardness scale $Q^2$.
The power is independent of the shape of the distribution amplitude(s), and is
for two-body scattering given by the dimensional scaling rule \cite{dsr}
\beq
\frac{d\sigma}{dt}(2\to 2) \propto \frac{f(t/s)}{t^{n-2}}  \label{scarul}
\eeq
where $n$ is the total number of elementary fields (quarks, gluons, photons)
that are involved in the scattering. It should be kept in mind that the
logarithmic scaling violations (\ref{qdep}) of the distribution amplitude(s)
as well as the running of $\as(Q^2)$ induces further $Q^2$-dependence which at
low scales can be appreciable.

The proton elastic and $p \to N^*$ transition form factors are in remarkably
good agreement with the scaling behavior of \eq{scarul} \cite{ss} (with the
notable exception of $p \to \Delta(1232)$ \cite{cm}). Dimensional scaling is
seen also in the $\gamma p \to \pi^+ n$ $90^\circ$ cross section \cite{gp90}
and in the deuteron elastic form factor \cite{fd}. Remarkably precocious
scaling is evidenced by $\sigma(\gamma d \to p n) \propto E_{CM}^{-22}$ at
$89^{\circ}$ for
$E_\gamma = 1 \ldots 4$ GeV \cite{bochna} (see Fig. 2).

\begin{figure}
\begin{center}
\mbox{\epsfig{file=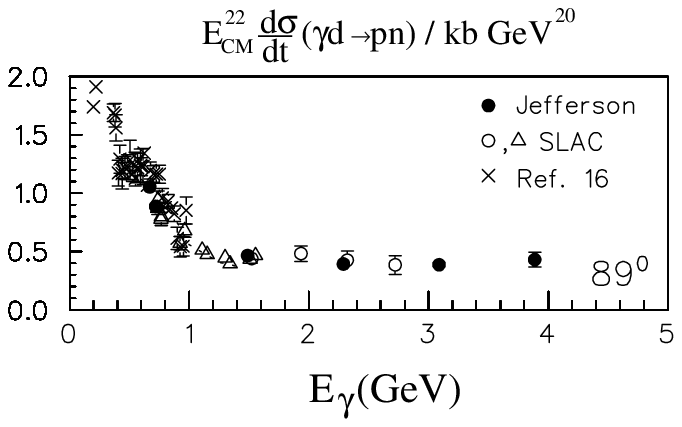,width=120mm}}
\end{center}
\captive{The $\gamma d \to pn$ cross section at $89^\circ$ multiplied by
$E_{CM}^{22}$ as a function of the photon beam energy [9].
\label{Fig2_Lund}}
\end{figure}

The early scaling observed in many exclusive processes is a pleasant surprise.
The overall momentum transfer is shared by many exchanges in the hard
subprocess $T_H$, implying that typical parton virtualities are quite low
\cite{ss}. Moreover, the cross sections involve high powers of $\as$,
suggesting considerable $Q^2$ dependence from a running coupling. This has been
interpreted \cite{freeze} as evidence that $\as$ `freezes' at low scales,
allowing PQCD to be applied and avoiding extra dependence on $Q^2$. On the
other hand, it has also been suggested that the scaling is only apparent, and
does not yet represent the true asymptotic behavior \cite{soft}.

A quantitative test of the exclusive PQCD formalism is provided by
measurements of the pion transition form factor $F_{\pi\gamma}$ in the process
$\gamma^* \gamma \to \pi^0$. Using the asymptotic form \cite{bl}
$\varphi_\pi^{as}(x) =\sqrt{3}f_\pi x(1-x)$ one predicts $F_{\pi\gamma} =
\sqrt{2} f_\pi/Q^2$ at large $Q^2$, where $f_\pi$ is the pion decay constant.
This prediction is in good agreement with the data \cite{ss}.

A direct measurement of the shape of hadron distribution amplitudes is
possible in diffractive processes. For example, in 
$\pi^- A \to jet_1(xp_\pi,k_\perp) + jet_2((1-x)p_\pi,-k_\perp)+A$ one expects
the longitudinal momentum sharing of the jets to reflect that of the pion
valence quark amplitude at a transverse scale $r_\perp \sim
1/k_\perp$ \cite{dd}. Preliminary data from the E791 experiment \cite{ashery}
using a 500 GeV $\pi^-$ beam on C and Pt targets indicates a nearly asymptotic
shape $\varphi_\pi(x,Q^2 \sim 10\ \gev^2) \sim x(1-x)$.
The nuclear target dependence of the cross section (integrated over the
diffractive peak), $\sigma \propto A^{1.6\pm 0.1}$, shows that the nucleus
is nearly transparent to the pion, which therefore must be in a configuration
of small transverse size.

\subsection{Skewed parton distributions}

Among more recent theoretical PQCD developments I would like to mention
the concept of Skewed Parton Distributions (SPD) \cite{spd}. These generalized
distributions incorporate properties of both the ordinary DIS parton
distributions and of the exclusive distribution amplitudes. For example, the
scattering amplitude of the process $\gamma^* p \to \pi^+ n$ (Fig. 3a)
factorizes in the standard Bjorken limit ($Q^2,\nu \to
\infty$ with $x_B=Q^2/2m\nu$ fixed for the virtual photon) into a hard (PQCD
calculable) upper vertex and an SPD (Fig. 3b).

\begin{figure}
\begin{center}
\mbox{\epsfig{file=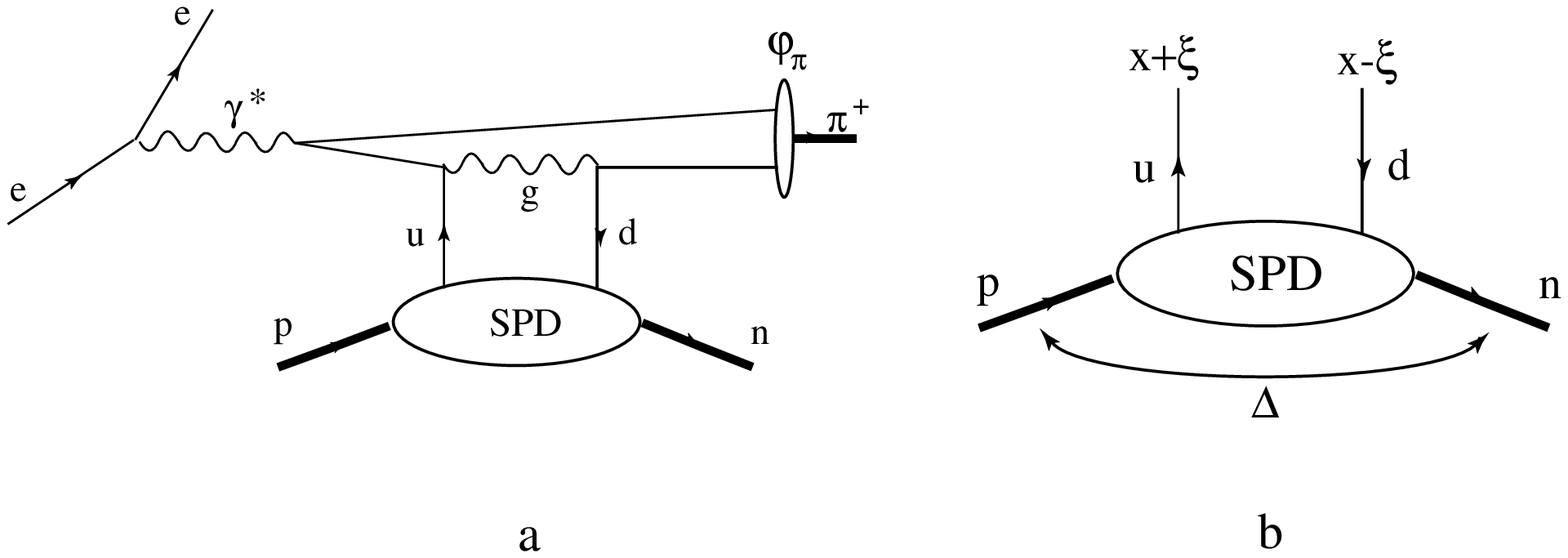,width=160mm}}
\end{center}
\captive{a. Lowest order contribution to the process $ep \to e\pi^+n$, in
the limit of large photon energy $\nu$ and virtuality $Q^2$ but a fixed
momentum transfer between the nucleons. b. A skewed parton distribution,
with the momentum fractions $x\pm\xi$ of the quarks and the momentum transfer
$\Delta$ between the nucleons indicated.
\label{Fig3_Lund}}
\end{figure}

The SPD differs from the standard forward distributions in that there is a
non-vanishing momentum transfer $\Delta$ from the initial to the final
nucleon. While the scattering can proceed with vanishing transverse exchange,
$\Delta_\perp=0$, the inelastic kinematics fixes a longitudinal momentum
fraction $2\xi \simeq x_B/(1-x_B/2)$. The longitudinal fraction $x$ of Fig. 3b
is integrated over in the scattering amplitude of Fig. 3a. In the region $x <
\xi$ the $d$-quark of Fig. 3b should be thought of as an outgoing $\bar d$
antiquark, \ie, the scattering occurs off a compact $q\bar q$ pair as in the
pion distribution amplitude.

Lorentz invariance allows 4 independent SPD's which depend on three
kinematic variables (in addition to the scale $Q^2$), often denoted
$H(x,\xi,\Delta^2),\ \tilde H,\ E$ and $\tilde E$ (precise definitions may be
found in Ref. \cite{vgg}). In the forward direction the diagonal SPD's (for
which $d, u$ in Fig. 3b are identical quarks $q$) are related to the standard
DIS quark structure functions,
\beqa
H^q(x,0,0) &=& q(x) \nonumber\\
\tilde H^q(x,0,0) &=& \Delta q(x) \label{stddis}
\eeqa
where $\Delta q(x)$ is the helicity distribution. Furthermore, there are sum
rules relating the SPD's to the nucleon elastic form factors (and hence,
via \eq{epep}, to the nucleon distribution amplitude),
\beqa
\int_{-1}^{+1}dx\,H^q(x,\xi,\Delta^2) &=& F_1^q(\Delta^2) \nonumber\\
\int_{-1}^{+1}dx\,E^q(x,\xi,\Delta^2) &=& F_2^q(\Delta^2) \label{formf}
\eeqa
where it may be noted that the integral is predicted to be independent of
$\xi$.

The first moment of the quark SPD's at $\Delta^2=0$ gives the total angular
momentum carried by the quark,
\beq
J^q = \int_{-1}^{+1}dx\,\left[H^q(x,\xi,0)+E^q(x,\xi,0)\right] =
\half\Delta\Sigma + L^q \label{qangmom}
\eeq
Since the quark spin $\Delta\Sigma/2$ is measured in polarized DIS, this
allows the contribution of the orbital angular momentum
$L^q$ to be determined. The total angular momentum carried by gluons can
analogously be determined from a relation like \eq{qangmom} for $H^g,E^g$.
The nucleon angular momentum can then be expressed as $J^q+J^g=1/2$.
It should be noted, however, that $\Delta^2=0$ lies outside the physical
region. Thus an extrapolation is needed to use measured data in the sum
rule (\ref{qangmom}).

It should be clear from the above that the skewed parton distributions
offer qualitatively new information on hadron structure. They are difficult to
measure due to the integral over $x$ in the expression for the scattering
amplitude (which must be squared to give a measured cross section).
Their phenomenology is at its infancy -- only the future and dedicated
facilities like ELFE can tell what they look like.

\section{ELFE Options}

The exclusive processes to be measured by ELFE have cross sections that
decrease rapidly with the photon virtuality $Q^2$ and hadron momentum transfer
$t$. For example, $d\sigma/dQ^2(ep \to ep) \propto Q^{-12}$. This puts
stringent demands on the performance of the accelerator.
\begin{itemize}
\item{\bf High luminosity.} Detectors with an open ($4\pi$) geometry are
estimated to be able to handle ${\cal L} \sim 10^{35}{\rm cm}^{-2}{\rm
s}^{-1}$, while
${\cal L} \sim 10^{38}{\rm cm}^{-2}{\rm s}^{-1}$ should be available for
spectrometers. This will allow measurements in the range $Q^2,|t| \lsim 20\
\gev^2$.
\item{\bf Energy 10\ldots 30 GeV.} Higher energies would not significantly
extend the momentum transfer range due to limitations in luminosity. ELFE will
be a high $x_B$ machine, which can reach into the cumulative $x_B>1$ region of
nuclei.
\item{\bf Good energy resolution.} A beam energy spread $\Delta E/E \lsim
10^{-3}$ is required to signal particles that escape direct detection.
\item{\bf Continuous beam.} Reconstruction of multi-particle final
states requires that consequtive events are separated in time. The
duty factor of the beam should be in excess of 50\%.
\item{\bf Polarization.} Measurements of helicity amplitudes and tests of
helicity conservation as predicted by PQCD require a polarized beam
($P_e>60\%$).
\end{itemize}

The first ELFE design was based on a recirculating beam \cite{elf1}. Later it
was realized \cite{elf2} that ELFE could be built in conjunction with the
TESLA $e^+e^-$ linear collider project at DESY, by using the HERA ring as a
stretcher for achieving a high duty factor. This ELFE@DESY option is
contingent upon an approval of TESLA. The time-scale is rather long, with
operation not foreseeable before 2010.

Recently, the possibility of using the RF cavities which are becoming
available at CERN at the closure of LEP has been considered. A conceptual
design report for ELFE@CERN is being prepared \cite{elf3}, which features a
recirculating accelerator in the North Area with 7 passes that could achieve
electron energies up to 25 GeV. The construction time would be on the order of
6 years after approval.

Should the European options not be realized, there is on a longer time scale
(ca. 2015) the possibility that the CEBAF accelerator at Jefferson Lab is
upgraded to 24 GeV. This would require a major reconstruction of the
machine. The long range plans of Jlab depend on the results of the
present upgrade program, which aims at achieving 12 GeV electron beams
by 2006.

\section{Concluding Remarks}

Many particle physicists believe that the
frontier of fundamental physics is synonymous with the frontier of energy, and
concerns physics beyond the Standard Model. Here I have advocated a frontier
related to the structure of hadrons and the physics of QCD. Basic
questions which are addressed at the QCD frontier include the treatment
of relativistic bound states and the physics of confinement.

Hadron physics is a largely blank page in our QCD book. The proton structure
functions (\ie, single parton distributions) only give a first glimpse at
proton structure. Fortunately, important progress is being made in widening
our understanding. The properties of perturbative QCD allow factorization
to be established for a variety of exclusive and semi-exclusive processes.
This makes new aspects of hadron wave functions accessible to measurement.
Experimental results show precocious scaling, hinting at a simple underlying
picture. The situation is reminiscent of the SLAC DIS measurements 30 years
ago.

The surprising simplicity of experimental results, both inclusive and
exclusive, is leading to a {\em faith transition} in QCD \cite{dokshitzer}:
The theory appears to be weakly coupled at long distance! There may be no
transition from a perturbative to a non-perturbative regime. As $Q^2 \to 0$,
the strong coupling freezes at a moderate value $\as(0)/\pi= 0.14 \ldots
0.17$. While this is still speculative, the mere possibility that we can use
the powerful tools of PQCD in the confinement regime is daunting. By taking
into account the `condensates' of quarks and gluons in the QCD ground state one
generates a PQCD series which is formally equivalent to the standard one,
but which may be able to describe the long-distance dynamics of QCD
\cite{softpqcd}.

Hadron physics is privileged in addressing fundamental questions through
an interplay between experiment and theory. The experiments are demanding, but
there seems little doubt that we have the technology to construct a facility
like ELFE. The importance of accurate data on exclusive processes for a variety
of beams and targets should also be kept in mind. Just as for hard inclusive
scattering, one can only trust QCD factorization provided all measurements of
hadron wave functions are self-consistent. New experimental results
will stimulate theoretical studies and {\em vice versa}. The door is
opening for progress in our experimental, theoretical and conceptual
understanding of hadrons.

\vskip 5mm\noindent
{\bf Acknowledgements.} I would like to thank the organizers for inviting me
to this interesting cross-disciplinary meeting covering physics at high energy
and synchrotron radiation facilities. I am grateful for a long and fruitful
collaboration on these topics with Stan Brodsky and for numerous interactions
with colleagues in the EU network HaPHEEP. This work was supported in part by
the EU/TMR contract EBR FMRX-CT96-0008.

\end{document}